\documentclass[floats,twocolumn,pra,footinbib]{revtex4}
\usepackage{eurosym}
\usepackage{graphics}
\usepackage[thmmarks]{ntheorem}
\usepackage{graphicx}
\usepackage{algorithm}
\usepackage{algpseudocode}
\usepackage{amsmath}
\usepackage{amsfonts}
\usepackage{amssymb}
\usepackage{float}
\usepackage{longtable}
\usepackage{flushend}
\usepackage{epsfig}
\usepackage{latexsym}
\usepackage{theorem}
\usepackage{bbm}
\usepackage{bm}
\usepackage{psfrag}
\usepackage{eurosym}
\usepackage[utf8]{inputenc}
\usepackage[T1]{fontenc}
\usepackage{physics}%
\usepackage{color}
\usepackage{multirow}
\usepackage{float}
\usepackage{soul}
\usepackage{hyperref}
\usepackage{mathrsfs}
\usepackage{cancel}
\hypersetup{
    colorlinks=true,
    linkcolor=blue,
    filecolor=magenta,      
    urlcolor=cyan,
}

\DeclareMathOperator{\diag}{diag}

\begin{document}
\title{Data post-processing for the one-way heterodyne protocol\\under composable finite-size security}
\author{Alexander G. Mountogiannakis}
\author{Panagiotis Papanastasiou}
\author{Stefano Pirandola}
\affiliation{Department of Computer Science, University of York, York YO10 5GH, United Kingdom}

\begin{abstract}
The performance of a practical continuous-variable (CV) quantum key distribution (QKD) protocol depends significantly, apart from the loss and noise of the quantum channel, on the post-processing steps which lead to the extraction of the final secret key. A critical step is the reconciliation process, especially when one assumes finite-size effects in a composable framework. Here, we focus on the Gaussian-modulated coherent-state protocol with heterodyne detection in a high signal-to-noise ratio regime. We simulate the quantum communication process and we postprocess the output data by applying parameter estimation, error correction (using high-rate, non-binary low-density parity-check codes), and privacy amplification. This allows us to study the performance for practical implementations of the protocol and optimize the parameters connected to the steps above. We also present an associated Python library performing the steps above.
\end{abstract}
\maketitle

\section{Introduction}
Based on physical laws and not on computational assumptions, quantum key distribution (QKD) ensures the creation of long secret keys between two distant authenticated parties, which can be later used for the exchange of symmetrically encrypted secret messages~\cite{revQKD}. In particular, the parties can trace any eavesdropper's action on their communication over the intermediate (insecure) quantum channel that links them. According to Heisenberg's principle, any attempt of the eavesdropper to interact with the traveling quantum signals leaves a trace~\cite{clone1}. Through this trace, the parties can quantify the leaked amount of information and compress their exchanged data appropriately, in order to decouple the eavesdropper from the final secret key. 

Traditionally, the parties exchange quantum states described by discrete degrees of freedom, such as the polarization of a photon~\cite{BB84}. Such schemes are called discrete-variable (DV) QKD protocols and their security has been studied copiously~\cite{revQKD,Hanzo}. More recently, quantum systems described by continuous degrees of freedom have also been studied and continuous-variable (CV) QKD protocols~\cite{GG02,noswitch} have emerged as alternatives to standard schemes. These degrees of freedom are observables such as the position and momentum of the electromagnetic field~\cite{revQKD,Gaussian_rev}. 

A great advantage of CV-QKD is that the current telecommunications infrastructure is capable of handling the preparation, exchange and detection of the corresponding quantum signals. Thus, it provides a cost-effective and practical solution, when compared with DV-QKD. Such protocols also provide high key rates over metropolitan-area distances~\cite{CVMDI}, with values approaching the theoretical limit of the secret key capacity, also known as the repeaterless PLOB bound~\cite{PLOB}. Lately, they have surpassed their previous performance in terms of achievable distances, which are now comparable to these of DV-QKD protocols~\cite{LeoEXP,LeoCodes}.

In practical applications, where the finite-size effects~\cite{UsenkoFNSZ} are important and the parties should take into account composable security terms~\cite{freeSPACE,QKDlevels}, the protocol performance declines. Therefore, the optimization of the protocol parameters becomes an important aspect in CV-QKD~\cite{QKD_SIM}. 
In particular, it is also important to optimize the procedure of data postprocessing, which is made up of various parts: parameter estimation (PE), raw key creation and error correction (EC), and  privacy amplification (PA). According to the composable framework, all these steps have associated error parameters that quantify the probability of failure for each process. These parameters are then combined into a final epsilon parameter that identifies the overall level of security provided by the protocol. 

In this work, we focus on the heterodyne protocol with Gaussian modulation of coherent states~\cite{noswitch} in a high signal-to-noise ratio regime. We simulate the quantum communication process and then postprocess the generated data
via PE, EC and PA by means of a dedicated Python library~\cite{Github}. In this way, we can evaluate the performance of this protocol, when it is deployed in realistic conditions and employed for high-speed quantum-secure communications
at relatively short ranges.
 
This is the summary of the manuscript. In Sec.~\ref{sec:CV-QKD with heterodyne detection}, we present the protocol and the calculation of its asymptotic key rate. 
In Sec.~\ref{sec:Composable key rate},  after the simulation of the quantum communication, we connect all the relevant parameters describing all the details of the post-processing steps with the composable key rate. 
In Sec.~\ref{sec:Sim}, we present the simulation specifications of the classical postprocessing while, in Sec.~\ref{sec:Results}, we comment and illustrate the results of our investigation. Finally, Sec.~\ref{sec:con} is for conclusions.
\section{\label{sec:CV-QKD with heterodyne detection}Review of CV-QKD with heterodyne detection}
\subsection{Protocol}
Alice draws samples from the variable $x$, which follows a normal distribution with zero mean and variance $\sigma_x^2=\mu-1$, i.e., described by the density function
\begin{equation}
    p(x)=(2\pi\sigma^2_x)^{-\frac{1}{2}}\exp\left[-\frac{x^2}{2 \sigma_x^2}\right].
\end{equation}
We denote the samples with $[x]_i$, where $i=1,\dots, 2N$.  Then  she groups them in instances $[\mathbf{x}]_j=([Q_x]_j,[P_x]_j)=([x]_{2j-1},[x]_{2j})$ for $j=1\dots N$ and encodes them in $n_\text{bks}$ blocks of coherent (signal) states $|\alpha_j\rangle$, where $\alpha_j=([x]_{2j-1}+\mathrm{i}[x]_{2j})/2$. We say that the block size is $N$. Note that we adopt the notation of Ref.~\cite[Sec.II]{Gaussian_rev} for the quadrature operators $(\hat{Q},\hat{P})$ so that $[\hat{Q},\hat{P}]=2\mathrm{i}$ and the vacuum noise variance is equal to $1$.

The coherent states travel to Bob through an optical fiber with length $L$ and loss rate $A_L$. This is simulated by a thermal-loss channel with transmissivity $T=10^{-\frac{A_L L}{10}}$ and $\bar{n}$ environmental photons. This is equivalent to assuming a beam splitter with transmissivity $T$ mixing the traveling mode $A$ from Alice with a mode $E$ of the environment in a thermal state with variance $\omega=2\bar{n}+1$. Then one may assume the dilation of the environmental state into a two-mode squeezed-vacuum (TMSV) state $\Phi_{Ee}$ held by the eavesdropper, Eve. This is a zero-mean Gaussian state with covariance matrix (CM)
\begin{equation}
\mathbf{V}_{Ee}(\omega)=
\begin{pmatrix}
\omega \mathbf{I}&\sqrt{\omega^2-1}\mathbf{Z}\\
\sqrt{\omega^2-1}\mathbf{Z}&\omega \mathbf{I}
\end{pmatrix},
\end{equation}
where $\mathbf{I}:=\diag\{1,1\}$ and $\mathbf{Z}:=\diag\{1,-1\}$. Note that this is the so-called ``entangling cloner'' attack; it is the most realistic form of a collective Gaussian attack~\cite{collectiveG}. 

Bob then decodes the signal states by applying a heterodyne measurement on the arriving mode $B$. The heterodyne measurement is performed by mixing $B$ with a vacuum mode through a balanced beam splitter. Then, homodyne measurement is applied to each output of the beam splitter, with respect to a different (conjugate) quadrature. In that way, Bob obtains to output instances $[\mathbf{y}]_j=([Q_y]_{j},[P_y]_{j})=([y]_{2j-1},[y]_{2j})$ for the  $j$th coherent state that encodes Alice's instances $[\mathbf{x}]_j$~\footnote{In the entanglement based (EB) representation EB, $P_y$  is anti-correlated with $P_x$ but with the same absolute correlation value as the $Q$-quadrature. As one may see from Eq.~\eqref{eq:mutual_rho}, the mutual information is not affected by the sign of the correlations. In this case, the parties can  change the sign of $P_y$ so that the two quadratures can form a single variable $x$  for Alice (and $y$ for Bob) with the same properties.}.

More precisely, Bob's detectors are characterized by efficiency $\eta$ and electronic noise $\upsilon_{\text{el}}$. As a result, the decoding variable is connected to the encoding one via
\begin{equation}\label{eq:input-output}
    y=\sqrt{T\eta}x+z,
\end{equation}
where $z$ is a Gaussian noise variable characterizing Bob's output. It has zero mean and variance equal to 
\begin{equation}
    \sigma_z^2=2+\upsilon_{\text{el}}+\Xi,
\end{equation}
where $\Xi=\eta T \xi $ is the variance of the channel's noise and 
\begin{equation}
    \xi:=\frac{1-T}{T}(\omega-1),
\end{equation}
is the channel's excess noise.
\subsection{Asymptotic rate}
In the asymptotic regime, where $N$ is large, one may calculate the mutual information between the parties theoretically based on the input-output relation of Eq.~(\ref{eq:input-output}). The variance of Bob's variable is given by
\begin{equation}
    \sigma_y^2=\eta T \sigma_x^2+\sigma_z^2,
\end{equation}
while the corresponding conditional variance on the input $x$ is given by
\begin{equation}
    \sigma_{y|x}^2= \sigma_y^2(\sigma_x^2=0)=\sigma_z^2.
\end{equation}
Because the variables $x$ and $y$ are Gaussian, the mutual information is given by
\begin{align}\label{eq:mutual_info}
    I(\mathbf{x}:\mathbf{y})=2I(x:y)=\log_2\left(\frac{\sigma_y^2}{\sigma_z^2}\right)=\log_2\left(1+\text{SNR}\right),
\end{align}
with
\begin{equation}\label{eq:SNR}
\text{SNR}=\frac{\sigma_x^2}{\sigma_z^2/(\eta T)}.
\end{equation}
Asymptotically, the maximum shared information between the parties is quantified by Eq.~(\ref{eq:mutual_info}). This is true when the efficiency of the reconciliation between the parties is ideal: In a practical reverse reconciliation scenario, Bob helps Alice's guessing of his outcome by publicly revealing more information than needed. This extra information leads to $\beta  I(\mathbf{x}:\mathbf{y})$, where $\beta \in (0,1]$ is known as the reconciliation efficiency.

In line with the definition of collective Gaussian attack, we assume that Eve stores her modes (after Gaussian interaction with the signal modes) into a quantum memory which she can optimally measure at the end of all quantum communication between the parties. The parties are able to quantify the maximum possible amount of leaked information by virtue of the Holevo bound. 
This is computed from the von Neumann entropies $S(\rho_{E'e})$ and $S(\rho_{E'e|y})$, in turn calculated from the joint CM of Bob and Eve. In particular, we have that 
\begin{equation}\label{eq:CM_beE}
 \mathbf{V}_{BeE'}=\begin{pmatrix}
 b \mathbf{I}&\gamma \mathbf{Z}&\theta \mathbf{I}\\
 \gamma \mathbf{Z}&\omega \mathbf{I} & \psi \mathbf{Z}\\
 \theta \mathbf{I}&\psi \mathbf{Z}&\phi \mathbf{I}
 \end{pmatrix}
\end{equation}
with 
\begin{align}
b:=& \eta T  (\mu+\xi)+1-T \eta+\upsilon_\text{el},\\
\gamma:=&\sqrt{\eta (1-T) (\omega^2-1)},\\
\theta:=&\sqrt{\eta T (1-T)}(\omega-\mu),\\
\psi:=&\sqrt{T(\omega^2-1)},\\
\phi:=&T\omega+(1-T)\mu.
\end{align}
By tracing out mode $B$ from Eq.~(\ref{eq:CM_beE}), we obtain $\mathbf{V}_{eE'}$. 

Then, by setting
\begin{equation}
 \mathbf{C}=\begin{pmatrix}
 \gamma \mathbf{Z}&\theta \mathbf{I}
 \end{pmatrix},   
\end{equation}
and applying the formula for the heterodyne measurement~\cite{Gaussian_rev}, we obtain Eve's conditional CM
\begin{align}
\mathbf{V}_{eE^{\prime}|\mathbf{y}} 
&  =\mathbf{V}_{eE^{\prime}}-(b+1)^{-1}\mathbf{C}^{T}\mathbf{C}\\
&  =\left(
\begin{array}
[c]{cc}%
\omega\mathbf{I} & \psi\mathbf{Z}\\
\psi\mathbf{Z} & \phi \mathbf{I}%
\end{array}
\right)  -(b+1)^{-1}\left(
\begin{array}
[c]{cc}%
\gamma^{2} \mathbf{I} & \gamma \theta \mathbf{Z}\\
\gamma \theta \mathbf{Z} & \theta^{2} \mathbf{I}%
\end{array}
\right).
\end{align}
Then we may write the Holevo information as
\begin{align}\label{eq:holevo}
    \chi(E:\mathbf{y})=&S(\rho_{E'e})-S(\rho_{E'e|\mathbf{y}})\\
    &=h(\nu_+)+h(\nu_-)-h(\tilde{\nu}_+)-h(\tilde{\nu}_-),
\end{align}
where
\begin{equation}
    h(\nu):=\frac{\nu+1}{2}\log_2\frac{\nu+1}{2}-\frac{\nu-1}{2}\log_2\frac{\nu-1}{2}
\end{equation}
and $\{\nu_\pm\}$, $\{\tilde{\nu}_\pm\}$ are the symplectic spectra of $\mathbf{V}_{eE'}$ and $\mathbf{V}_{eE^{\prime}|\mathbf{y}}$ respectively.
Finally, the asymptotic secret key rate will be given by
\begin{align}\label{eq:AsyRate}
R_\text{asy}=\beta I(\mathbf{x}:\mathbf{y})-\chi(E:\mathbf{y})\\
=R(\beta,\mu,\eta,\upsilon_{\text{el}},T,\xi).
\end{align}
\section{\label{sec:Composable key rate} Composable key rate}
In this section, we describe the effects of PE, EC and PA on the final secret key rate in the finite-size regime where these steps cannot be considered ideal but may have outputs that fail to have the desired properties with a small probability, i.e., $\tilde{\epsilon}_\text{PE}$, $\epsilon_\text{cor}$ and $\epsilon_\text{sec}$ respectively.


\subsection{Channel parameter estimation}
For each block, the parties randomly choose $m$ instances $[\mathbf{x}]_j$ and $[\mathbf{y}]_j$ and broadcast them through the public channel.
The parties use the corresponding samples $[x]_i$ and $[y]_i$ from all the blocks assuming a stable channel~\footnote{We assume  that experimentally the coherent state preparation can be done quite fast. In this regime, in the time interval $\Delta t$ it is feasible to be produced $N n_\text{bks}$ states while the transmissivity of the channel can still be considered constant.}. Based on these $M=2mn_\text{bks}$ samples they define the maximum likelihood estimators (MLEs) 
\begin{equation}
\widehat{T}=\frac{1}{\eta(\sigma_x^2)^2}\left(\widehat{C}_{xy}\right)^2
\end{equation}
with
\begin{equation}
\widehat{C}_{xy}=\frac{1}{M}\sum_{k=1}^{M}[x]_k [y]_k
\end{equation}
and 
\begin{equation}
\widehat{\Xi}=\widehat{\sigma_z^2}-\upsilon_\text{el}-2~~\text{for}~~\widehat{\sigma_z^2}=\frac{1}{M}\sum_{k=1}^{M}\left([y]_k-\sqrt{\eta \widehat{T}}[x]_k\right)^2.
\end{equation}
Based on a theoretical analysis as in Ref.~\cite{QKD_SIM}, one can find worst-case values for the above estimators so as to bound Eve's accessible information. These are given by
\begin{equation}\label{eq:wcest}
T_M=\widehat{T}-w\sigma_{\widehat{T}},~~~ \Xi_M=\widehat{\Xi}+w\sigma_{\widehat{\Xi}}
\end{equation}
with 
\begin{equation}
\sigma^2_{\widehat{T}}=\frac{2}{M}\widehat{T}^2\left(2+\frac{\widehat{\sigma_z^2}}{\eta\widehat{T}\sigma_x^2}\right),~~~\sigma^2_{\widehat{\Xi}}=\frac{(\widehat{\sigma_z^2})^2}{M}
\end{equation}
and
\begin{equation}
    w=\sqrt{2}\text{erf}^{-1}(1-\epsilon_\text{PE})
\end{equation}
where $\epsilon_\text{PE}$ is the failure probability of $T_M$ and $\Xi_M$ to be the worst-case scenario values for bounding Eve's information. The overall failure probability (combining the two events) is
\begin{equation}\label{eq:overall_ePE}
2\epsilon_\text{PE}(1-\epsilon_\text{PE})+\epsilon_\text{PE}^2\leq2\epsilon_\text{PE}.
\end{equation}
Taking into consideration the previous parameters, we can derive the asymptotic rate after parameter estimation
\begin{equation}\label{eq:Mrate}
R_M=\beta I(\mathbf{x}:\mathbf{y})|_{\widehat{T},\widehat{\Xi}}-\chi(E:\mathbf{y})|_{T_M,\Xi_M}.
\end{equation}
From the formula in Ref.~\cite[Eq.~(8.56)]{CoverThomas}, the mutual information of the variables $x$ and $y$
\begin{equation}\label{eq:mutual_rho}
    I(x:y)=\frac{1}{2} \log_2 \left[ 1+\text{SNR}\right]=\frac{1}{2}\log_2 \left [(1-\rho^2)^{-1}\right]
\end{equation}
is connected with their correlation
\begin{equation}
\rho:=\frac{\mathbb{E}(xy)}{\sigma_x\sigma_y}=\sqrt{\frac{\text{SNR}}{1+\text{SNR}}}.
\end{equation}

Therefore, one may derive the estimator for the correlation between the variables by replacing the MLEs of transmissivity and noise in Eq.~\eqref{eq:SNR}, namely, 
\begin{align}\label{eq:practical cor}
   \widehat{\rho}
   &=\sqrt{\frac{\sigma_x^2}{\sigma_x^2+\widehat{\sigma}_z^2/(\eta\widehat{T})}}.
\end{align}
Note that this is going to be used later in the \emph{a priori} probabilities of Sec.~\ref{sec:Data rec}. 

\subsection{Error correction\label{sec:Data rec}}
Given that $M$ signal states have been processed through PE ($m$ per block), only  $n=N-m$ per block are available for secret key extraction. More specifically, before the step of PA, Alice and Bob need to reconcile over their raw data strings ($2 n n_\text{bks}$ samples), in order to end up with identical strings up to some small error probability $\epsilon_\text{cor}$. The preprocessing of EC contains the steps of normalization, discretization and splitting. During EC, blocks of data with errors that cannot be corrected get discarded with probability $1-p_\text{EC}$. The remaining blocks are combined into a large string, which is used as input to the next step of PA.

\subsubsection{Normalization}
Alice and Bob concatenate the $n=N-m$ samples from each block in order to calculate the estimated variancess~\footnote{We assume here that the variables $x$ and $y$ have a zero mean value. Alternatively, the parties  subtract the mean value $\bar{x}$ and $\bar{y}$ of $x$ and $y$ respectively from their instances to create updated centered variables $x\leftarrow x-\bar{x}$ and $y\leftarrow y-\bar{y}$. Then the formulas for estimating the variance keep the same form as in Eq.~\eqref{eq:classical_variance_est}.}
\begin{equation}\label{eq:classical_variance_est}
    \widehat{\sigma_x^2}=\frac{1}{n_\text{ent}}\sum_{k=1}^{n_\text{ent}}[x]_k^2,~~\widehat{\sigma_y^2}=\frac{1}{n_\text{ent}}\sum_{i=1}^{n_\text{ent}}[y]_k^2
\end{equation}
for $n_\text{ent}=2nn_\text{bks}$. Then they divide the values $[x]_i$ by the standard deviation $\widehat{\sigma_x}=\sqrt{ \widehat{\sigma_x^2}}$ and the values $[y]_i$ by the other standard deviation $\widehat{\sigma_y}=\sqrt{ \widehat{\sigma_y^2}}$, therefore creating the normalized samples $[X]_i$ and $[Y]_i$, following a bivariate normal distribution with CM
\begin{equation}
    \Sigma_{XY}=\begin{pmatrix}
        1&\rho\\
        \rho&1
    \end{pmatrix}.
\end{equation}
In terms of a practical calculation from the data, we use $\widehat{\rho}$ from Eq.~\eqref{eq:practical cor}.

\subsubsection{Discretization}
Bob maps each of the  samples $[Y]_i$ into a number $\mathsf{l}=0,\dots,2^p-1$ for $p$ an integer and obtains the corresponding samples $[\mathsf{l}]_i$ according to a one-dimensional lattice with cut-off parameter $\alpha$ and step $\delta=2\alpha2^{-p}$. More specifically, he computes $\delta$-size intervals, i.e., bins, $[a_\mathsf{l},b_\mathsf{l})$ with boundary points  given according to 
\begin{equation}\label{eq:left border}
  a_\mathsf{l}=\begin{cases}
 -\infty~~~~~~~~\text{for}~~\mathsf{l}=0,\\\\
  - \alpha +\mathsf{l} \delta~~~\text{for}~~\mathsf{l}>0,
  \end{cases}
 \end{equation}
 and
\begin{equation}\label{eq:right border}
  b_\mathsf{l}=\begin{cases}
  -\alpha + (\mathsf{l}+1) \delta~~~\text{for}~\mathsf{l}<2^p-1,\\\\
  \infty~~~~~~~~~~~~~~~~~~~\text{for}~\mathsf{l}=2^p-1.
  \end{cases}
 \end{equation}

Alice computes the conditional probability of  the value  $\mathsf{l}$ given  the value $X$. She then obtains
\begin{equation}\label{eq:cond_prob}
  P(\mathsf{l}|X)=\frac{1}{2}\text{erf}\left(\frac{b_\mathsf{l}-\widehat{\rho}X}{\sqrt{2(1-\widehat{\rho}^2)}}\right)-\frac{1}{2}\text{erf}\left(\frac{a_\mathsf{l}-\widehat{\rho}X}{\sqrt{2(1-\widehat{\rho}^2)}}\right).
\end{equation}

\subsubsection{Splitting}
Bob then splits each sampled symbol $[\mathsf{l}]_i$ into top $[\overline{\mathsf{l}}]_i$ and bottom $[\underline{\mathsf{l}}]_i$ symbols. More specifically, he chooses numbers $q$ and $d$, such that $q+d=p$, and breaks each $p$-ary symbol $\mathsf{l}$ into a $q$-ary symbol $\overline{\mathsf{l}}$ and $d$-ary symbol $\underline{\mathsf{l}}$  respectively according to the rule:
\begin{equation}\label{eq:splitting}
\mathsf{l}=\overline{\mathsf{l}}2^d+\underline{\mathsf{l}}.
\end{equation}
Alice then calculates the probability for a specific top symbol $\overline{\mathsf{l}}=0,\dots,2^q-1$, given its bottom counterpart $\underline{\mathsf{l}}=0,\dots,2^d-1$ and the  variable $X$. She then obtains
\begin{equation}\label{eq:aprioriprob}
P(\overline{\mathsf{l}}|X\underline{\mathsf{l}})=\frac{P(\overline{\mathsf{l}},\underline{\mathsf{l}}|X)}{\sum_{\underline{\mathsf{l}}}P(\overline{\mathsf{l}},\underline{\mathsf{l}}|X)},
\end{equation}
where $P(\overline{\mathsf{l}},\underline{\mathsf{l}}|X)$ is given by Eq.~\eqref{eq:cond_prob}.

\subsubsection{LDPC encoding and decoding}
Let us assume the reverse reconciliation scenario, where Alice guesses Bob's sequence. Ideally, Bob's sequence is described by the continuous variables $\mathbf{y}$. Given that Alice knows the variable $\mathbf{x}$ correlated with $\mathbf{y}$ by the quantum channel and that Bob's entropy is $H(\mathbf{y})$, Bob needs to send $H(\mathbf{y}|\mathbf{x})$ bits of information  through a public channel, if we wanted  Alice's accessible information to be  equal to the mutual information
\begin{equation}\label{eq:mi_accessible}
    I(\mathbf{x}:\mathbf{y})=H(\mathbf{y})-H(\mathbf{y}|\mathbf{x}).
\end{equation}
Note that the previous entropic quantities refer to the average number of bits exchanged per signal state (i.e. including both quadratures). Let us assume the variable $\mathsf{l}$ to be the discretized version of $Y$. After the previous classical postprocessing, it holds that
\begin{align}
H(\mathbf{y})=&H(Q_y,P_y)=2H(y)\geq 2H(Y)\label{eq:equality_lq_lp_0}\\
&\geq 2H(\mathsf{l})=H(Q_\mathsf{l})+H(P_\mathsf{l})=H(Q_\mathsf{l},P_\mathsf{l})=H(l)\label{eq:equality_lq_lp}
\end{align}
where  the variables  $Q_l$ and $P_l$ correspond to samples with odd and even indexes respectively and
\begin{equation}\label{eq:bi-directional map} 
l=Q_\mathsf{l} 2^p+P_\mathsf{l}
\end{equation}
is a bidirectional mapping. 
Note that Eq.~\eqref{eq:equality_lq_lp_0} is true, because $Q_y$ and $P_y$ are independent (the same holds, later, for  $Q_\mathsf{l}$ and $P_\mathsf{l}$ as different samples of an i.i.d. variable). Furthermore, we compare the (differential) entropy of two Gaussian variables, $y$ with variance $\sigma_y^2$ and $Y$ with unit variance  as the normalized version of $y$, which is dependent only on the variances of the two variables~\cite[Th.~17.2.3]{CoverThomas}. For passing from Eq.~\eqref{eq:equality_lq_lp_0} to Eq.~\eqref{eq:equality_lq_lp} one may use the joint entropy of $Y$ and $l$ ~\cite{entropic} and observe that $l$ is a deterministic outcome of $Y$ (while the opposite is not true). The last equation in  Eq.~\eqref{eq:equality_lq_lp} holds because the mapping in Eq.~\eqref{eq:bi-directional map} is bidirectional~\cite{one-to-one}. In particular, the parties estimate $H(l)$ through  $H(\mathsf{l})$. To increase the accuracy of the estimation result, the parties estimate the previous quantity including all the samples $[\mathsf{l}]_i$ from all the $n_\text{bks}$ blocks. Then the estimate is given by
\begin{equation}
    \widehat{H}(\mathsf{l})=-\sum_\mathsf{l} \nu_\mathsf{l}\log_2\nu_\mathsf{l}
\end{equation}
where $\nu_\mathsf{l}$ is the frequency of the value $\mathsf{l}$ in the samples $[\mathsf{l}]_i$ from all the $n_\text{bks}$ blocks. For this estimator the following inequality is true~\cite{kontogiannis}:
\begin{equation}\label{eq:wcent}
    H(\mathsf{l})\geq \widehat{H}(\mathsf{l})-\delta_\text{ent}
\end{equation}
where 
\begin{equation}\label{eq:bound_entropy}
    \delta_\text{ent}=\log_2(n_\text{ent})\sqrt{\frac{2\log(2/\epsilon_\text{ent})}{n_\text{ent}}}
\end{equation}
up to an error probability $\epsilon_\text{ent}$.

In a realistic situation, Alice is guessing a sequence of discrete symbols and Bob sends information through the public channel equal to $n^{-1}\text{leak}_\text{EC}\geq H(\mathbf{y}|\mathbf{x})$. 
The top samples are sent to Alice through the public channel, encoded by a regular LDPC with code rate $R_\text{code}$, constant for any length $2n$. For the LDPC encoding, the $[\overline{\mathsf{l}}]_i$ are considered to be elements of the Galois field $\mathcal{GF}(2^q)$. Bob  builds a $c\times 2 n$ sparse parity check matrix $\mathbf{H}$ such that $c/(2 n)=1-R_\text{code}$~\cite{QKD_SIM}. He then calculates the syndrome $\mathsf{l}_\text{syn}^{c}= \mathbf{H}\overline{\mathsf{l}}^{2n}$ for each block and sends it to Alice, while the bottom sequence is publicly revealed. 
In other words, Bob is sending at most 
\begin{equation}
(k/(2n)) q + d = (1-R_\text{code}) q+d=-R_\text{code} q+p
\end{equation}
bits per sample $[\mathsf{l}]_i$. It is clear that $d$ should be as small as possible, yet not negligible, in order for this reconciliation scheme  to succeed.  This bounds the leakage term per signal state:
\begin{equation}\label{eq:leakage_bound}
n^{-1}\text{leak}_\text{EC}\leq 2( -R_\text{code} q+p).
\end{equation}

\textit{\subparagraph{Remark~1} Note that in the case of an active concatenation of the quadratures, i.e.  creating the variable $l$ of Eq.~\eqref{eq:bi-directional map}, the parties will have to perform error correction on $n$ symbols, which are described by $2p$ bits for each block. This will demand  a higher  value for $q$  approximately raised to $q^\prime \simeq 2 q$. Subsequently, this will increase crucially the requirements for computational resources, in order to achieve the same speed for EC.}

By replacing $H(\mathbf{y})$ with $H(l)$ from Eq.~\eqref{eq:equality_lq_lp} and $H(\mathbf{y}|\mathbf{x})$  with $n^{-1}\text{leak}_\text{EC}$  in~\eqref{eq:mi_accessible}, we obtain that
\begin{align}\label{eq:first_replacement}
I(\mathbf{x}:\mathbf{y})\geq H(l)-n^{-1}\text{leak}_\text{Ec}:=\beta I(\mathbf{x}:\mathbf{y}),
\end{align}
where $\beta$ is the reconciliation efficiency.
Finally, we consider the practical calculation of $\beta$. We first bound the term
\begin{equation}
    H(l)-n^{-1}\text{leak}_\text{Ec}\geq 2(\hat{H}(\mathsf{l})- \delta_\text{ent})-2(-R_\text{code}q+p)
\end{equation}
considering the estimation of $\hat{H}(\mathsf{l})$ and the bound for the $\text{leak}_\text{EC}$. Then we also assume the value of the mutual information for the estimated channel parameters $I(x:y)_{\hat{T},\hat{\Xi}}$. Therefore one obtains the practical reconciliation efficiency
\begin{equation}\label{eq:recEF}
    \hat{\beta}=2\frac{\widehat{H}(\mathsf{l})+R_\text{code} q-p- \delta_\text{ent}}{I(x:y)|_{\widehat{T},\widehat{\Xi}}}.
\end{equation}
\subparagraph{Remark 2}\textit{The previous equation can be written in terms of the SNR as
\begin{equation}
 \hat{\beta}=\frac{\widehat{H}(\mathsf{l})+R_\text{code} q-p-\delta_\text{ent}}{\frac{1}{2}\log_2(1+\widehat{\text{SNR}})}.
\end{equation}
This equation is equal to the corresponding one for the homodyne protocol (see \cite[Eq.~(56) and~(76)]{QKD_SIM}). In fact, it returns the same results, given that the SNR is the same for both protocols (different combination of transmissivity, excess noise and classical modulation variance.) }

Then one may set  $\hat{\beta} I(\mathbf{x}:\mathbf{y})|_{\widehat{T},\widehat{\Xi}}:= 2[\widehat{H}(\mathsf{l})+R_\text{code} q-p-\delta_\text{ent}]$ in Eq.~\eqref{eq:Mrate} to obtain
\begin{equation}
R_M^\text{EC}=2[\widehat{H}(\mathsf{l})+R_\text{code} q-p-\delta_\text{ent}]-\chi(E:\mathbf{y})|_{T_M,\Xi_M}.
\label{eq:Mrate2}\end{equation}
We also obtain
\begin{align}\label{eq:LDPCcode}
R_\text{code}&=\left(\hat{\beta} I(x:y)|_{\widehat{T},\widehat{\Xi}}/2+p+\delta_\text{ent}-\widehat{H}(\mathsf{l})\right)q^{-1}\\
&=\left((\beta/2) \log_2(1+\eta \widehat{T}\sigma^2_x/\widehat{\sigma^2_z})+p+\delta_\text{ent}-\widehat{H}(\mathsf{l})\right)q^{-1}.
\end{align}
Alice then uses the probabilities of Eq.~(\ref{eq:aprioriprob}) to initialize a sum-product algorithm~\cite{QKD_SIM} with a maximum number of iterations $\text{iter}_\text{max}$. During every iteration, the algorithm finds a sequence $\hat{\mathsf{l}}^{2n}$ that is optimal for the given likelihood, calculates its syndrome, and compares it with   $\mathsf{l}_\text{sd}^{c}$. If the syndromes are equal, the specific block qualifies for the verification step. If they are not equal, the algorithm continues to the next iteration. In case the maximum number of iterations $\text{iter}_\text{max}$ is exceeded, the given sequence is discarded, along with its associated bottom counterpart.

\subsubsection{Verification}
The strings $\hat{\mathsf{l}}^{2n}$ and $\overline{\mathsf{l}}^{2n}$ with the same syndrome are turned into binary strings $\hat{\mathsf{l}}^{2n}_\text{bin}$ and $\overline{\mathsf{l}}_\text{bin}^{2n}$ respectively over which the parties calculate hashes of $\lceil-\log_2 \epsilon_\text{cor}\rceil$ bits (For more details on the calculation of the hashes see Ref.~\cite{QKD_SIM}). The parties check their hashes and if they are equal, they are certain that their sequences agree with a probability $1-\epsilon_\text{cor}$ for a very small $\epsilon_\text{cor}$. Then, they concatenate the binary version of the bottom sequence $\underline{\mathsf{l}}_\text{bin}^{2n}$ to $\hat{\mathsf{l}}_\text{bin}^{2n}$ and $\overline{\mathsf{l}}_\text{bin}^{2n}$, creating the sequences
\begin{equation}\label{eq:PA_sequencies}
\widehat{S}=\hat{\mathsf{l}}_\text{bin}^{2n}\underline{\mathsf{l}}_\text{bin}^{2n}~~~\text{and}~~~S=\overline{\mathsf{l}}_\text{bin}^{2n}\underline{\mathsf{l}}_\text{bin}^{2n}.
\end{equation}
If the hashes do not agree, the strings $\hat{\mathsf{l}}_\text{bin}^{2n}$, $\overline{\mathsf{l}}_\text{bin}^{2n}$ and $\underline{\mathsf{l}}^{2n}$ are discarded. From the ratio of the sequences that pass to the PA over the total number $n_\text{bks}$ of sequences, one calculates the probability $p_\text{EC}$ of EC.


\subsection{\label{sec:PA}Privacy amplification}
Privacy amplification is the final step that creates the secret key from the raw shared data. The parties start with two different sequences of $n_\text{bks}$ blocks, each block with $2N$ samples. After postprocessing, these are reduced to two indistinguishable (with probability $1-\epsilon_\text{EC}$) binary sequences, that consist of $p_\text{EC}n_\text{bks}$ blocks, each block carrying $2np$ bits (see Eq.~\eqref{eq:PA_sequencies}). 

The parties then decide to further compress their data in order to prevent Eve from having any knowledge of their bit sequences. To do so, they concatenate their previous sequences into large ones $\mathbf{S}\simeq \widehat{\mathbf{S}}$ containing $\tilde{n}:=2p_\text{EC}n_\text{bks}np$ bits
and compress them using a universal hashing: They apply a Toeplitz matrix $\mathbf{T}_{r,\tilde{n}}$ to their sequences (see more details in Ref.~\cite{QKD_SIM}) in order to extract the secret key 
\begin{equation}
    \mathbf{K}=\mathbf{T}_{r,\tilde{n}}\mathbf{S}\simeq \mathbf{T}_{r,\tilde{n}}\widehat{\mathbf{S}}
\end{equation}
which has length $r=p_\text{EC}n_\text{bks}n\tilde{R}$ where $\tilde{R}$ is the composable key rate. The latter takes into account any small distance of the practical protocol from  an ideal one.  More specifically, each of the processes of PE and EC has small failure probabilities $\tilde{\epsilon}_\text{PE}$ and  $\epsilon_\text{cor}$. 

In $\tilde{\epsilon}_\text{PE}=2\epsilon_\text{PE}+\epsilon_\text{ent}$, we include the overall failure probability of PE (see Eq.~ Eq.~\eqref{eq:overall_ePE}) and the failure probability of bounding the Bob's variable entropy $\epsilon_\text{ent}$ (see Eq.~\eqref{eq:bound_entropy}). The PA procedure is characterized by the $\epsilon$-secrecy parameter, which quantifies the potential failure to completely exclude Eve from obtaining information about the key with probability $\epsilon_\text{sec}$. The latter can be broken in two separate parameters: the smoothing parameter $\epsilon_\text{s}$ and the hashing parameter $\epsilon_\text{h}$, which yield $\epsilon_\text{sec}=\epsilon_\text{s}+\epsilon_\text{h}$. The composition of all these parameters (see also Eq.~\eqref{eq:compasable security}) defines the security parameter of the protocol
\begin{equation}
    \epsilon=p_\text{EC}(2\epsilon_\text{PE}+\epsilon_\text{ent})+\epsilon_\text{cor}+\epsilon_\text{sec},
\end{equation}
with typical choice $\epsilon_\text{s}=\epsilon_\text{h}=\epsilon_\text{cor}=\epsilon_\text{PE}=\epsilon_\text{ent}=2^{-32}\simeq 2.3 \times 10^{-10}$, so that for any value of  $p_\text{EC}$ we have $\epsilon \lesssim 10^{-9}$.
Finally, the secret key rate of the protocol, in terms of bits per channel use, takes the form~\cite{freeSPACE}%
\begin{equation}
R=\frac{np_{\text{EC}}}{N}\tilde{R},~\tilde{R}:=\left(R_{M}^{\mathrm{EC}}-\frac{\Delta_{\text{AEP}}}{\sqrt{n}}+\frac{\Theta}{n}\right), \label{rateCOMPO}
\end{equation}
where $R_{M}^{\mathrm{EC}}$ is the rate of Eq.~(\ref{eq:Mrate2}) where $\beta$ is  replaced by  Eq.~(\ref{eq:recEF}) and the extra terms are~\cite{freeSPACE,QKDlevels}
\begin{align}
&  \Delta_{\text{AEP}}:=4\log_{2}\left(2^{p} + 2\right)  \sqrt{\log
_{2}\left(  \frac{18}{p_{\text{EC}}^{2}\epsilon_{\text{s}}^{4}}\right)
},\label{deltaAEPPP}\\
&  \Theta:=\log_{2}[p_{\text{EC}}(1-\epsilon_{\text{s}}^{2}/3)]+2\log
_{2}\sqrt{2}\varepsilon_{\text{h}}. \label{bigOMEGA}
\end{align}
Note that the discretization bits $p$ appear in $\Delta_{\text{AEP}}$ providing a total dimension of $2^{2p}$ per symbol (see Appendix~\ref{app:Virtual concatenation of the conjugate quadrature variables}). One may also compare the previous rate with the corresponding theoretical rate 
\begin{equation}
R_\text{theo}=\frac{np_\text{EC}}{N}R^\star,~~ R^\star:=\left(\bar{R}_M-\frac{\Delta_{\text{AEP}}}{\sqrt{n}}+\frac{\Theta}{n}\right)
\end{equation}
where $\bar{R}_M$ has been computed  based on the initial values of the channel parameters used to produce the simulation data. In fact, one may replace in Eq.~\eqref{eq:Mrate} the mean value of the estimators and obtain
\begin{equation}
\bar{R}_M=\beta I(\mathbf{x}:\mathbf{y})|_{\bar{T},\bar{\Xi}}-\chi(E:\mathbf{y})|_{\bar{T}_M,\bar{\Xi}_M},
\end{equation}
where the following substitutions have been made:
\begin{align}
\widehat{T}&\leftarrow\bar{T}:=\mathbb{E}(T)\simeq T+\mathcal{O}(1/M), \\
\widehat{\Xi}&\leftarrow\bar{\Xi}:=\mathbb{E}(\widehat{\Xi})\simeq\Xi
\end{align}
and
\begin{align}
    T_M&\leftarrow\bar{T}_M:=T-w\sigma_{T},\\
    \Xi_M&\leftarrow \bar{\Xi}_M:=\Xi+w\sigma_{\Xi}
\end{align}
with
\begin{equation}
\sigma^2_{T}=\frac{2}{M}T^2\left(2+\frac{\sigma_z^2}{\eta T\sigma_x^2}\right),~~~\sigma^2_{\Xi}=\frac{(\sigma_z^2)^2}{M}.
\end{equation}
On the other hand, in the previous rate the parameters $p_{EC}$ and $\beta$ have been calculated through the simulation; in fact, they are known after EC (see Fig.~\ref{fig:R_vs_N}).  
\section{\label{sec:Sim}Simulation}
Here, we summarize the steps of the heterodyne protocol simulation taking into account the finite-size effects in a composable framework. Our approach follows steps similar to those of the homodyne protocol in Ref.~\cite{QKD_SIM}. Despite the fact that the simulation steps of the two protocols are quite similar, we want here to present a summary of the heterodyne protocol for the sake of completeness. We also have the opportunity to clarify some differences between the two simulations  because of the use of different formulas.  


\begin{description}
\item[Preparation:] Alice encodes $2Nn_{\text{bks}}$ samples $[x]_{i}$ of the generic variable $x\sim\mathcal{N}(0,\mu-1)$ on the two conjugate quadratures of $Nn_{\text{bks}}$ coherent states. In particular, the samples with an odd index will be encoded in the $Q$-quadrature of the $Nn_{\text{bks}}$ coherent states, while those with an even index will be encoded in the $P$-quadrature of the coherent states.

\item[Measurement:] During the decoding step, Bob obtains $2Nn_{\text{bks}}$ output samples
$[y]_i$ of $y=\sqrt{\eta T}x+z$ according to the propagation of the channel and the projection based on the heterodyne measurement.

\item[Public declaration:] Bob chooses an average of $m$ instances from each block and reveals them and their positions through the public channel. In each block, an average of $n$ instances are left for key generation.


\item[Estimators:]The parties use  $ M=2 m n_\text{bks}$ samples to define MLEs $\widehat{T}$ and $\widehat{\Xi}$  for $T$ and $\Xi$, respectively.
Then, by setting a PE error $\epsilon_\text{PE}$, they can calculate the values $T_M$ and $\Xi_M$ for the channel parameters. These values constitute the worst-case scenario assumption on the collected data with probability $1-\epsilon_\text{PE}$.

\item[Normalization:] 
Alice and Bob normalize the samples $[x]_i$ and $[y]_i$  dividing them by their practical standard deviations $\widehat{\sigma_x}$ and $\widehat{\sigma_y}$, creating the samples $[X]_i$ and $[Y]_i$, respectively.
These variables now follow a standard normal distribution. 

\item[Discretization:] Bob maps every sample $[Y]_i$ into a number $\mathsf{l}=0,\dots, 2^p-1$. To do so, he creates a one-dimensional lattice  for the values of the standard normal distribution  with cut-off parameter $\alpha$ and step $\delta=\alpha2^{1-p}$. Alice calculates the conditional probabilities $P(\mathsf{l}|X)$.

\item[Splitting:]  Bob splits $[\mathsf{l}]_i$  into two samples, $[\overline{\mathsf{l}}]_i$ and $[\underline{\mathsf{l}}]_i$. In fact, he derives a top $q$-ary symbol and a bottom $d$-ary symbol from $\mathsf{l}$  according to
\begin{equation}
\mathsf{l}=\overline{\mathsf{l}}2^d+\underline{\mathsf{l}}.
\end{equation}
Finally, Alice computes the \emph{a priori} probabilities $P(\overline{\mathsf{l}}|X,\underline{\mathsf{l}})$.

\item[LDPC encoding:] 
From the estimated SNR
and the practical Shannon entropy $\widehat{H}(\mathbf{\mathsf{l}})$,
the parties calculate the rate of the LDPC code according to Eq.~\eqref{eq:LDPCcode}.
Bob then calculates the $c\times 2n$ parity-check matrix $\mathbf{H}$, for $c=2n(1-R_\text{code})$, with entries in $\mathcal{GF}(p)$, and computes the syndrome $\mathsf{l}^c_\text{sd} =\mathbf{H}\overline{\mathsf{l}}^{2n}$. Bob sends the syndromes and the bottom sequences  $\underline{\mathsf{l}}^{2n}$ to Alice through the public channel for every block.

\item[LDPC decoding:] Alice updates the likelihood function  (initially equal to the product of the \emph{a priori} probabilities, see Eq.~\eqref{eq:aprioriprob}) using a sum-product algorithm. This update takes place  with respect to the syndrome $\mathsf{l}^c_\text{sd}$. After every iteration of the algorithm, the output likelihood function becomes the input for the next iteration. At the same time, Alice finds $\hat{\mathsf{l}}^{2n}$ that maximizes the updated likelihood function. She then compares the syndrome of $\hat{\mathsf{l}}^{2n}$ with $\mathsf{l}^c_\text{sd}$, and if they are equal, the algorithm terminates and gives as output the string $\hat{\mathsf{l}}^{2n}$, i.e. Alice's guess for $\overline{\mathsf{l}}^{2n}$. Otherwise, the algorithm continues to the next iteration until a maximum number of iterations $\text{iter}_\text{max}$ is reached. If the algorithm is not able to determine a guess after  $\text{iter}_\text{max}$, the given block is discarded and does not participate in the final key.

\item[Verification:] Alice's guess $\hat{\mathsf{l}}^{2n}$ and Bob's sequence $\overline{\mathsf{l}}^{2n}$ are converted  into binary sequences $\hat{\mathsf{l}}^{2n}_\text{bin}$ and $\overline{\mathsf{l}}^{2n}_\text{bin}$ respectively. Then both parties compute hashes of $\lceil-\log_2\epsilon_\text{cor}\rceil$ bits over their sequences. Bob discloses his hash and Alice compares it with hers. In case they are identical, they concatenate their string with the binary version of the bottom string $\underline{\mathsf{l}}_\text{bin}$ and obtain the strings 
\begin{equation}
    \widehat{S}:=\hat{\mathsf{l}}^{2n}_\text{bin}\underline{\mathsf{l}}^{2n}_\text{bin}\simeq S:=\overline{\mathsf{l}}^{2n}_\text{bin}\underline{\mathsf{l}}^{2n}_\text{bin},
\end{equation}
respectively, which are further promoted to the privacy amplification step (PA). Otherwise, the strings $\hat{\mathsf{l}}^{2n}_\text{bin}$, $\overline{\mathsf{l}}^{2n}_\text{bin}$ and $\underline{\mathsf{l}}^{2n}_\text{bin}$ are discarded, and the given block does not participate in the final key.


\item[Privacy amplification:] The parties concatenate the strings $\widehat{S}$ and $S$ from every block into long binary sequences $\widehat{\mathbf{S}}$ and $\mathbf{S}$ of $\tilde{n}=p_\text{EC}n_\text{bks}n2p$ bits. Given a level of secrecy $\epsilon_\text{sec}$, the parties calculate the composable rate $\tilde{R}$ and compress the sequences $\widehat{\mathbf{S}} \simeq \mathbf{S}$ with the use of a Toeplitz matrix $\mathbf{T}_{r,\tilde{n}}$ into the final secret key $\mathbf{K}$ of length $r:=p_\text{EC}n_\text{bks}n\tilde{R}$.

\end{description}

\section{\label{sec:Results}Results}

Since the protocol of this paper is better suited to short-range distances, only distances up to $5$km are examined. Consequently, the SNR of the performed simulations is relatively high and takes values from $\sim 6$ to $10$. Two features are considered essential in achieving a positive composable secret key rate $R$. The first is having a sufficient number of total key generation states $nn_\text{bks}$. The second is the choice of the reconciliation efficiency, which must be large enough to obtain a high rate but small enough to comfortably execute error correction. A large number of total key generation states will also lead to a better value for the reconciliation efficiency. This connection is provided by the presence of $\delta_\text{ent}$ term in Eq. (\ref{eq:recEF}), which becomes smaller as the number of states increases.

\begin{figure}[t]
    \centering
    \includegraphics[width=0.48\textwidth]{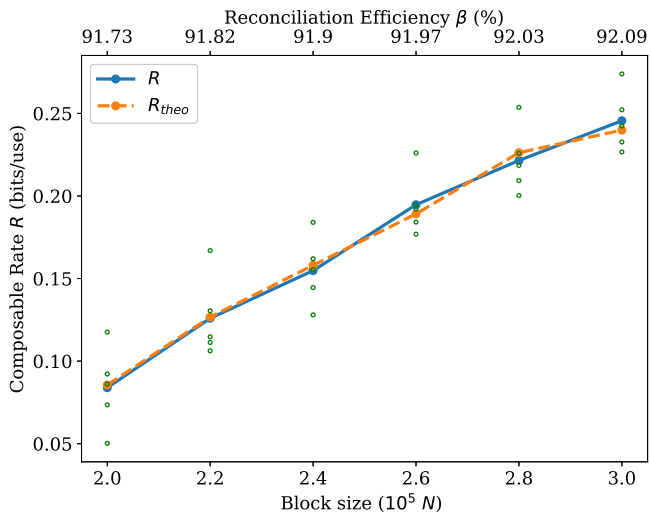}
    \caption{Composable secret key rate $R$ (bits/use) versus the block size $N$ for $\text{SNR}=10$. We compare the rate of Eq.~(\ref{rateCOMPO}) from five simulations (green points) and their average (blue solid line) with the theoretical rate $R_\text{theo}$ in Sec.~\ref{sec:PA} (orange dashed line), where the theoretical guesses for $\tilde{\beta}$ and $\tilde{p}_{\mathrm{EC}}$ are chosen compatibly with the simulations. For every simulation, $\tilde{p}_{\mathrm{EC}}=p_{\mathrm{EC}}$ has been set. All simulations have achieved $p_\text{EC} \geq 0.9$. The step of $N$ is $20000$. The values of the reconciliation efficiency $\beta$ are shown on the top axis and are chosen so as to produce $R_\text{code} \approx 0.846$. See Table~\ref{res_param1} for the list of input parameters used in the simulations.}
    \label{fig:R_vs_N}
\end{figure}

\begin{figure}[t]
    \centering
    \includegraphics[width=0.48\textwidth]{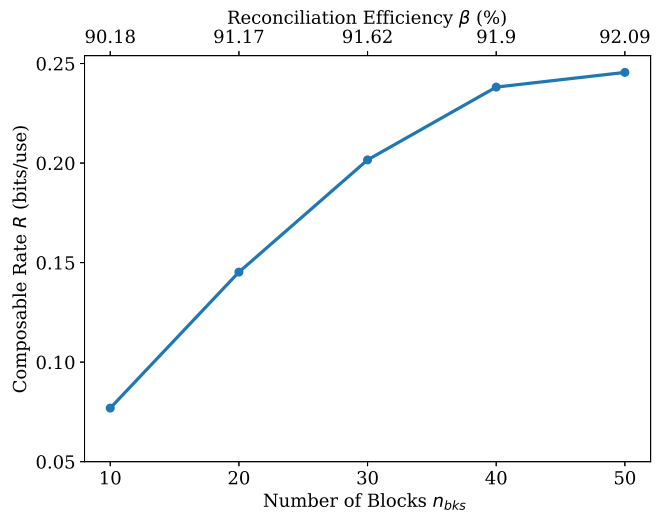}
    \caption{Composable secret key rate $R$ (bits/use) versus the number of blocks $n_\text{bks}$ for $\text{SNR}=10$. The step of $n_\text{bks}$ is $10$. The individual block size is fixed and equal to $N=3 \times 10^5$. Every point represents the average value of $R$, which is obtained after 5 simulations. All simulations have achieved $p_\text{EC} \geq 0.9$. The values of the reconciliation efficiency $\beta$ are shown on the top axis and are chosen so as to produce $R_\text{code} \approx 0.846$. See Table~\ref{res_param1} for the list of input parameters used in the simulations.}
    \label{fig:R_vs_nbks}
\end{figure}

A demonstration of sample parameters, that achieve a positive composable key rate and how this rate varies, according to changes in the block size $N$ and the number of blocks $n_\text{bks}$, is shown in Figs.~\ref{fig:R_vs_N} and \ref{fig:R_vs_nbks} respectively. Alice's signal variance $\mu$ is tuned so as to produce a rather high signal-to-noise ratio ($\text{SNR}=10$). It is observed in Fig. \ref{fig:R_vs_N} that a block size of at least $2\times10^5$ is needed. Additionally, Fig. \ref{fig:R_vs_nbks} shows that it is possible to yield higher key rates with fewer total states, if an adequately large block size $N$ is specified.

\begin{figure}[t]
    \centering
    \includegraphics[width=0.48\textwidth]{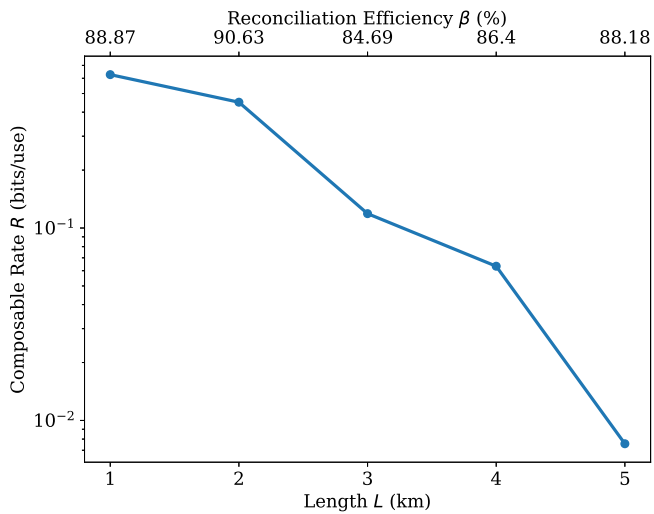}
    \caption{Composable secret key rate $R$ (bits/use) versus the channel length $L$ (km). Here we use $N=3.6 \times 10^{5}$. Every point represents the average value of $R$, which is obtained after 5 simulations. All simulations have achieved $p_\text{EC} \geq 0.9$. The values of the reconciliation efficiency $\beta$ are shown on the top axis. Other parameters are taken as in Table~\ref{res_param1}.}
    \label{fig:R_vs_L}
\end{figure}

Fig. \ref{fig:R_vs_L} portrays the composable rate $R$ versus distance $L$, expressed in km of standard optical fiber. Here, the $\text{SNR}$ varies from $5.732$ to $6.887$. For this simulation, the discretization bits value was set to $p=6$, in order to reach farther distances. A higher value for $p$ would severely limit the protocol's ability to achieve a positive $R$ at distances larger than $3$ km.

\begin{table*}[t]
\vspace{0.1cm}
\begin{tabular}
[c]{|c|c|c|c|c|c|c|}\hline
Parameter & Symbol & Value (Fig.~\ref{fig:R_vs_N}) & Value (Fig.~\ref{fig:R_vs_nbks}) & Value (Fig.~\ref{fig:R_vs_L}) & Value (Fig.~\ref{fig:R_vs_xi}) & Value (Fig.~\ref{fig:R_vs_SNR})  \\ \hline
Channel length & $L$ & $3$ & $3$ & variable & $4$ & $5$  \\
Attenuation & $A$ & $0.2$ & $0.2$ & $0.2$ & $0.2$ & $0.2$ \\
Excess noise & $\xi$ & $0.01$ & $0.01$ & $0.01$ & variable & $0.01$ \\
Setup efficiency & $\eta$ & $0.85$ & $0.85$ & $0.8$ & $0.85$ & $0.85$ \\
Electronic noise & $\upsilon_{\text{el}}$ & $0.1$ & $0.1$ & $0.1$ & $0.05$ & $0.1$ \\
Number of blocks & $n_{\text{bks}}$ & $50$ & variable & $50$ & $50$ & $50$  \\
Block size & $N$ & variable & $3 \times 10^{5}$ & $3.6 \times 10^{5}$ & $4.5 \times 10^{5}$ & $4 \times 10^{5}$  \\
Number of PE runs & $M$ & $0.1n_{\text{bks}}N$ & $0.1n_{\text{bks}}N$ & $0.1n_{\text{bks}}N$ & $0.1n_{\text{bks}}N$ & $0.1n_{\text{bks}}N$  \\
Discretization bits & $p$ & $7$ & $7$ & $6$ & $6$ & variable  \\
Most significant bits & $q$ & $4$ & $4$ & $4$ & $4$ & $4$ \\
Phase-space cut-off & $\alpha$ & $7$ & $7$ & $7$ & $7$ & $7$  \\
Max EC iterations & $\text{iter}_{\text{max}}$ & $100$ & $100$ & $150$ & $100$ & $150$  \\
Epsilon parameters & $\epsilon_{\text{\text{PE}, s, h, \text{corr}}}$ & $2^{-32}$ & $2^{-32}$ & $2^{-32}$ & $2^{-32}$ & $2^{-32}$ \\
Signal variance & $\mu$ & $\approx29.46$ & $\approx29.46$ & $20$ & $25$ & variable  \\ \hline
\end{tabular}
\caption{The input parameters for the simulations.}\label{res_param1}
\end{table*}

\begin{figure}[t]
    \centering
    \includegraphics[width=0.48\textwidth]{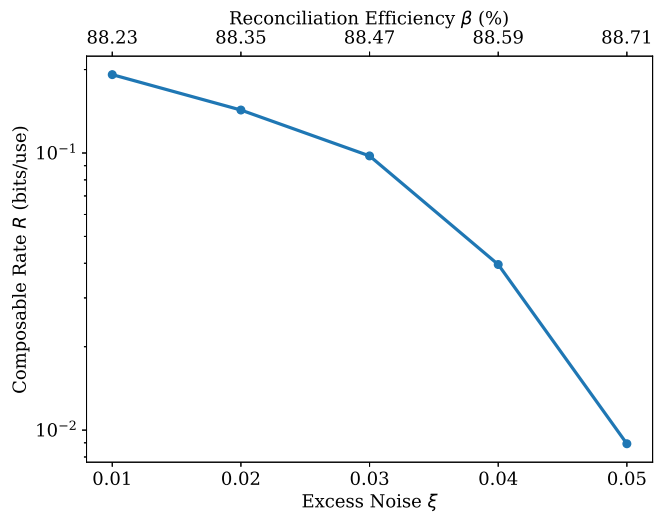}
    \caption{Composable secret key rate $R$ (bits/use) versus the excess noise $\xi$. Every point represents the average value of $R$, which is obtained after 5 simulations. Here we use $N=4.5 \times 10^{5}$ and $n_\text{bks}=50$. The values of the reconciliation efficiency $\beta$ for the heterodyne protocol simulations are chosen so as to produce $R_\text{code} \approx 0.8$. Other parameters are taken as in Table~\ref{res_param1}.}
    \label{fig:R_vs_xi}
\end{figure}

Fig. \ref{fig:R_vs_xi} presents an estimate of the maximum tolerable excess noise $\xi$. The variables used here produce an SNR of somewhat above $8$. While the decrease of the SNR is fairly small as the excess noise increases, the composable rate declines rapidly. In addition, the reconciliation efficiencies used here are in the range of $88.23$ - $88.71$. Such values provide efficient error correction but are not ideal for attaining a positive rate in the composable framework. Therefore, to achieve a positive rate at $\xi=0.05$, a large block size ($N=450000$) has to be used. 

\begin{figure}[!ht]
    \centering
    \includegraphics[width=0.48\textwidth]{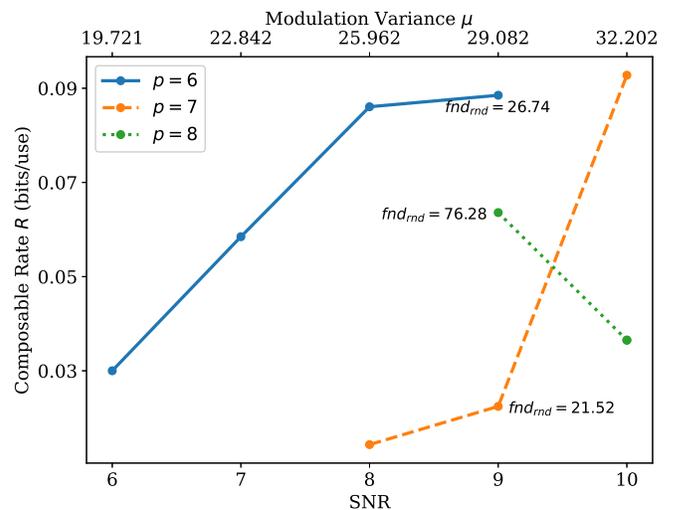}
    \caption{Composable secret key rate $R$ versus SNR for discretization bits $p=6$ (blue solid line), $p=7$ (orange dashed line) and $p=8$ (green dotted line). The chosen reconciliation efficiency $\beta$ for each value of the $\text{SNR}$ is shown in Table~\ref{recon_efficiencies_choice}. Every point represents the average value of $R$, which is obtained after 5 simulations. For SNR = 9, the average number of iterations $\text{fnd}_\text{rnd}$ needed to decode and verify a block is displayed for every point next to their respective points. The signal variance $\mu$ that was used to achieve the respective SNR is displayed on the top axis with an accuracy of 3 decimal digits. Other parameters are chosen as in Table~\ref{res_param1}.} 
    \label{fig:R_vs_SNR}
\end{figure}

Fig. \ref{fig:R_vs_SNR} describes the behaviour of the key rate against different SNR values, when the noise terms are fixed and the modulation variance is variable. If the same code rate is used, lower values of $p$ (at a fixed $q=4$), return higher rates for the corresponding $\text{SNR}$. It is possible for a higher $p$ value to yield a better composable rate than a smaller $p$, given that a larger code rate, and therefore larger reconciliation efficiency, is employed. An example is given by cases `a' and `b' of $\text{SNR}=9$, whose code rates and reconciliation efficiencies are shown in Table~\ref{recon_efficiencies_choice}. A combination of $p=8$ and $\beta=0.9301$ beats the combination of $p=7$ and $\beta=0.894$ in terms of the composable rate by a fairly large margin. However, the trade-off here is that the EC stage of the former combination requires plenty more iteration rounds, making the procedure more computationally expensive. 
Furthermore, for certain code rates, a minimum value for $p$ is required. Such an occasion is the `b' case of $\text{SNR}=9$, where error correction can only be achieved for $p=8$. Smaller values for $p$ would not be able to achieve error correction and, consequently, a positive rate.

\begin{table}[t]
\begin{tabular}
[c]{|l|l|l|l||l|l|}
\hline
$\text{SNR}$ & $\beta_{p=6}$ & $\beta_{p=7}$ & $\beta_{p=8}$  & $R_\text{code}$ & $d_{c}$ \\\hline
$6$ & $0.8651$  & & &    $0.75$                & $8$ \\
$7$ & $0.8836$  & & & $0.777$          & $9$ \\
$8$ & $0.8924$  & $0.8910$ & & $0.8$   & $10$ \\
$9_\text{a}$    & $0.8953$   & $0.8940$ &  & $0.818$ & $11$ \\
$9_\text{b}$ & &  & $0.9301$ & $0.833$ & $12$ \\
$10$           & $0.9244$  & $0.9231$ & $0.9229$ & $0.846$            & $13$ \\
\hline
\end{tabular}
\caption{The chosen reconciliation efficiency $\beta$ for each $\text{SNR}$ of Fig.~\ref{fig:R_vs_SNR}, together with its respective code rate $R_\text{code}$ and the row weight $d_{c}$ of the LDPC code. A missing value for the reconciliation efficiency implies that the returned composable key rate will most likely be negative under the specified values. The column weight $d_{v}$ remains constant and equal to $2$ for all simulations.} \label{recon_efficiencies_choice}
\end{table}

\section{\label{sec:con}Conclusion}

In this work, we  completely characterized the post-processing of data generated from a numerical simulation of the CV-QKD protocol, based on Gaussian modulation of coherent states and heterodyne detection. In particular, we designed the data post-processing accounting for the various composable finite-size terms arising from a realistic representation of the protocol.
Correspondingly, we provided a Python library for simulation, optimization and data post-processing specifically tailored for the considered heterodyne protocol. For future developments, one possible line of research is to extend these techniques to free-space scenarios~\cite{Zuo,freeSPACE} where fading effects are explicitly accounted in the protocol (e.g., via the use of pilots) and suitably mitigated via additional post-processing techniques.

\smallskip
\textbf{Acknowledgements}.--~A. M. was supported by the EPSRC via a Doctoral Training Partnership (Grant No. EP/R513386/1). P.P. was supported by the EPSRC via the UK Quantum Communications Hub (Grant No. EP/T001011/1). 



\appendix

\section{\label{app:Virtual concatenation of the conjugate quadrature variables}Virtual concatenation of the conjugate quadrature variables}
What we present here is a review and direct adaptation of the theory developed in Appendix G of Ref.~\cite{freeSPACE}.
Let us assume  Bob's measurement variables are $\mathbf{y}=(Q_y,P_y)$. Bob maps these variables to $\mathsf{\mathbf{l}}=(Q_\mathsf{l},P_\mathsf{l})$ via analog-to-digital conversion (ADC). Then, the output classical-quantum state (CQ) of Alice, Bob and Eve, after the collective attack will be given by a state in a tensor product form $\rho^{\otimes n}$, where the single copy state will be given by
\begin{align}\label{eq:two-register state}
    \rho=&\sum_{\mathsf{\mathbf{k}},\mathsf{\mathbf{l}}}p(\mathsf{\mathbf{k}},\mathsf{\mathbf{l}})|\mathsf{\mathbf{k}}\rangle_{R_A}\langle\mathsf{\mathbf{k}}|\notag 
    \otimes|\mathsf{\mathbf{l}}\rangle_{R_B}\langle\mathsf{\mathbf{l}}|  \otimes \rho_E(\mathsf{\mathbf{k}},\mathsf{\mathbf{l}})
\end{align}
where $R_A$ and $R_B$ are Alice's and Bob's classical raw-key registers, $\mathsf{\mathbf{k}}=(Q_k,P_k)$ is the corresponding discretized version of Alice's encoding variable and $p(\mathsf{\mathbf{k}},\mathsf{\mathbf{l}})$
is the joint probability of the discretized variables. 

The tensor product state can be then written as 
\begin{align}
  \rho^{\otimes n}&=\sum_{\mathsf{\mathbf{k}}^n,\mathsf{\mathbf{l}}^n}p(\mathsf{\mathbf{k}}^n,\mathsf{\mathbf{l}}^n)|\mathsf{\mathbf{k}}^n\rangle_{R_A^n}\langle\mathsf{\mathbf{k}}^n|\notag 
    \otimes|\mathsf{\mathbf{l}}^n\rangle_{R_B^n}\langle\mathsf{\mathbf{l}}^n|  \otimes \rho_E^{\otimes n}(\mathsf{\mathbf{k}}^n,\mathsf{\mathbf{l}}^n)\notag\\
    &=\sum_{\mathsf{k}^{2n},\mathsf{l}^{2n}}p(\mathsf{k}^{2n},\mathsf{l}^{2n})|\mathsf{k}^{2n}\rangle_{R_A^n}\langle\mathsf{k}^{2n}|\notag \\
    &~~~~~~\otimes|\mathsf{l}^{2n}\rangle_{R_B^n}\langle\mathsf{l}^{2n}|  \otimes \rho_E^{\otimes n}(\mathsf{k}^{2n},\mathsf{l}^{2n}).
\end{align}
Here, we replace the sequence  $\mathsf{\mathbf{l}}^{n}$ with the sequence $\mathsf{l}^{2n}$ so that each element $[\mathsf{l}]_{2j-1}$ corresponds to the element $[Q_\mathsf{l}]_j$ and each element $[\mathsf{l}]_{2j}$ to the element $[P_\mathsf{l}]_j$ for $j=1\dots n$.

In RR, Alice guesses Bob's sequence $\mathsf{l}^{2n}$ with $\tilde{\mathsf{l}}^{2n}$ using her corresponding sequence $\mathsf{k}^{2n}$ and $\text{leak}_\text{EC}$ bits of information from Bob. The parties publicly compare the two hashes of length $\lceil 1 - \log_2 \epsilon_\text{cor} \rceil $ computed from $\tilde{\mathsf{k}}^{2n}$ and $\mathsf{l}^{2n}$ respectively. If they are equal, the 
parties continue with the protocol with probability $p_\text{EC}$; otherwise they abort. This procedure is associated with a residual failure probability $\epsilon_\text{cor}$, which bounds the probability of the two sequences being different, even if their hashes coincide
\begin{equation}
    p_\text{EC} \text{Prob}(\tilde{\mathsf{l}}^{2n}\neq\mathsf{l}^{2n})\leq p_\text{EC} 2^{-\lceil1-\log_2\epsilon_\text{cor}\rceil}\leq\epsilon_\text{cor}.
\end{equation}

In turn, EC can be simulated by a projection $\Pi_\mathfrak{S}$ of Alice's and Bob's classical registers $R^n_A$ and $R^n_B$  onto a ``good'' set $\mathfrak{S}$ of sequences. With success probability
\begin{equation}
    p_\text{EC}=\text{Tr}(\Pi_\mathfrak{S}\rho^{\otimes n}).
\end{equation}
This quantum operation generates a classical-quantum state 
\begin{equation}
    \tilde{\rho}^n:= p_\text{EC}^{-1}\Pi_\mathfrak{S}\rho^{\otimes n}\Pi_\mathfrak{S}
\end{equation}
which is restricted to those sequences $\{\mathsf{k}^{2n},\mathsf{l}^{2n}\}$
that can be corrected, i.e., mapped to a successful pair $\{\tilde{\mathsf{l}}^{2n},\mathsf{l}^{2n}\}$.

The parties continue with the PA step with probability $p_\text{EC}$ and apply a two-way hash function over $\tilde{\rho}^n$ which outputs the PA state $\bar{\rho}^n$, i.e., $\rho^{\otimes n}\rightarrow \tilde{\rho}^n \rightarrow \bar{\rho}^n$, with the later approximating the ideal state (defined below)
\begin{equation}
p_\text{EC}D(\bar{\rho}^n,\rho_\text{id})\leq\epsilon_\text{sec}.
\end{equation}
In fact, Alice and Bob perform EC and PA over the state $\rho^{\otimes n}$, in order to approximate the $s_n$-bit ideal classical-quantum state
\begin{equation}\label{eq:ideal_state}
\rho_{\text{id}}:=2^{-s_{n}}
{\displaystyle\sum\limits_{z=0}^{2^{s_{n}}-1}}
\left\vert z\right\rangle _{R_A^{n}}\left\langle z\right\vert \otimes\left\vert
z\right\rangle _{R_B^{n}}\left\langle z\right\vert \otimes\rho_{E^{n}},
\end{equation}
with Alice's and Bob's classical registers completely decoupled from Eve and containing the same completely-random sequence $z$ with length $s_n$.
Using the triangle inequality, one obtains~\cite[Th.~4.1]{Portman}
\begin{equation}\label{eq:epsilon_def}
p_\text{EC}D(\tilde{\rho}^n,\rho_\text{id})\leq \epsilon:=\epsilon_\text{cor}+\epsilon_\text{sec}.
\end{equation}
The state $\bar{\rho}^n$ will contain $s_n$ bits of shared uniform randomness satisfying the direct leftover hash bound
\begin{equation}\label{eq:leftover hash bound}
    s_n\geq H^{\epsilon_\text{s}}_\text{min}(\mathsf{l}^{2n}|E^n)_{\tilde{\rho}^n}+2\log_2\sqrt{2}\epsilon_\text{h}-\text{leak}_\text{EC}.
\end{equation}

Here  $H^{\epsilon_\text{s}}_\text{min}(\mathsf{l}^{2n}|E^n)_{\tilde{\rho}^n}$ is the smooth min-entropy of Bob's sequence $\mathsf{l}^{2n}$ conditioned on Eve's system $E^n$ after EC, and the smoothing $\epsilon_\text{s}$ and hashing $\epsilon_\text{h}$ parameters satisfy
\begin{equation}
    \epsilon_\text{s}+\epsilon_\text{h}=\epsilon_\text{sec}.
\end{equation}
In Eq.~(\ref{eq:leftover hash bound}) we explicitly account for the bits leaked to Eve during
EC. In fact, one may write $s_{n}\geq H_{\text{min}%
}^{\epsilon_{\text{s}}}(\mathsf{l}^{2n}|E^{n}R)_{\tilde{\rho}^{n}}+2\log_{2}\sqrt
{2}\epsilon_{\text{h}}$ where $R$ is a register of dimension $d_{R}%
=2^{\mathrm{leak}_{\text{ec}}}$, while $E^{n}$ are the systems used by Eve
during the quantum communication. Then, the chain rule for the smooth-min
entropy leads to $H_{\text{min}}^{\epsilon_{\text{s}}}(\mathsf{l}^{2n}|E^{n}%
R)_{\tilde{\rho}^{n}}\geq H_{\text{min}}^{\epsilon_{\text{s}}}(\mathsf{l}^{2n}%
|E^{n})_{\tilde{\rho}^{n}}-\log_{2}d_{R}$. As we have seen earlier (see Eq.~\eqref{eq:leakage_bound}), in the proposed EC procedure, Bob sends to Alice $p-R_\text{code}q$ bits for each of the quadratures in a signal state. This allows us to bound the leakage term by
\begin{equation}
   \text{leak}_\text{EC}\leq  2n( -R_\text{code} q+p).
\end{equation}
We now use that the previous result is connected with the smooth min-entropy of $\rho^{\otimes n}$, which will later allow the AEP approximation. In fact, one can show that (see~\cite[Appendix~G2]{freeSPACE})
\begin{align}\label{eq: EC projection}
    H^{\epsilon_\text{s}}_\text{min}(\mathsf{l}^{2n}|E^n)_{\tilde{\rho}^n}\geq H^{p_\text{EC}\epsilon^2_\text{s}/3}_\text{min}(\mathsf{l}^{2n}|E^n)_{\rho^{\otimes n}}\notag\\+\log_2\left[p_\text{EC}(1-\epsilon^2_\text{s}/3)\right].
\end{align}

Let us assume that the parties concatenate their discretized values corresponding to the two quadrature variables of a single channel use according to the \emph{bidirectional mapping}:
\begin{equation}\label{eq:bidirmap}
    l=Q_\mathsf{l}2^p+P_\mathsf{l}.
\end{equation}
In that sense, instead of labeling the classical states as in Eq.~\eqref{eq:two-register state} by using the combination of two labels, each described by $p$ bits, we use one label described by $2p$ bits. Therefore, we have a classical mapping from a state $\rho^{\otimes  n}:=\rho^{\otimes  n}_{\mathsf{l}^{2n}}$ described 
by the sequence $\mathsf{l}^{2n}$ to the state 
\begin{equation}
    \rho^{\otimes n}_{l^{n}}\leftarrow \rho^{\otimes n}_{\mathsf{l}^{2n}}
\end{equation}
described by the sequence $l^n$. In Eq.~\eqref{eq: EC projection}, this implies the following relation for the smooth  min-entropy of the two states:
\begin{equation}
H^{p_\text{EC}\epsilon^2_\text{s}/3}_\text{min}(\mathsf{l}^{2n}|E^n)_{\rho^{\otimes n}_{\mathsf{l}^{2n}}}\geq H^{p_\text{EC}\epsilon^2_\text{s}/3}_\text{min}(l^n|E^n)_{\rho^{\otimes n}_{l^{n}}},
\end{equation}
where we use Appendix~\ref{app:entropy for calssical mapping}.

Then, from the AEP theorem, one  obtains
\begin{align}
H^{p_\text{EC}\epsilon^2_\text{s}/3}_\text{min}(l^n|E^n)_{\rho^{\otimes n}_{l^{n}}}\geq n H(l |E)_{\rho}-\sqrt{n}\Delta_\text{AEP}(p_\text{EC}\epsilon_\text{s}^2/3,|\mathcal{L}|),\label{eq:AEP}
\end{align}
where $H(l|E)_{\rho}$ is the conditional von Neumann entropy computed over the single-copy state $\rho$ (after applying the mapping of Eq.~\eqref{eq:bidirmap}) and
\begin{equation}
    \Delta_\text{AEP}(\epsilon_\text{s},|\mathcal{L}|)=4\log_2(\sqrt{|\mathcal{L}|}+2)\sqrt{\log_2(2/\epsilon_s^2)}
\end{equation}
with $|\mathcal{L}|$ being the cardinality of the discretized variable $l$, i.e., in our case $2^{2p}$.
By combining Eqs.~\eqref{eq:leftover hash bound},~\eqref{eq: EC projection} and ~\eqref{eq:AEP}, we  write the following lower bound
\begin{align}\label{eq:VN bound}
s_n&\geq nH(l|E)_{\rho}-\sqrt{n}\Delta_\text{AEP}(p_\text{EC}\epsilon_\text{s}^2/3,2^{2p})\notag\\+&\log_2\left[p_\text{EC}(1-\epsilon^2_\text{s}/3)\right]+2\log_2\sqrt{2}\epsilon_\text{h}-\text{leak}_\text{EC}.
\end{align}
Note that for the conditional entropy, we have
\begin{equation}
    H(l|E)_{\rho}=H(l)-\chi(l:E)_{\rho}
\end{equation}
where $H(l)$ is the Shannon entropy of $l$ and $\chi(E:l)_{\rho}$ is  Eve's Holevo bound with respect to $l$. In more detail, using the data processing inequality, we have

\begin{equation}
\chi(E:l)_{\rho}\leq \chi(E:Q_y,P_y)=\chi(E:\mathbf{y})
\end{equation}
where the latter term is calculated using Eq.~\eqref{eq:holevo}. Therefore we have
\begin{equation}\label{eq:rep_holevo}
     H(l|E)_{\rho} \geq H(l)-\chi(E:\mathbf{y})
\end{equation}
Furthermore, we may make the following replacement (see also Eq.~\eqref{eq:first_replacement})
\begin{equation}\label{eq:rep_MI}
H(l)-n^{-1}\text{leak}_\text{EC}=\beta I(\mathbf{x}:\mathbf{y})
\end{equation}
where $I(\mathbf{x}:\mathbf{y})$ is calculated from Eq.~\eqref{eq:mutual_info} and 
\begin{equation}
    \beta=\frac{H(l)-n^{-1}\text{leak}_\text{EC}}{I(\mathbf{x}:\mathbf{y})}
\end{equation}
is the reconciliation efficiency.

Replacing Eq.~\eqref{eq:rep_MI} and~\eqref{eq:rep_holevo} in~\eqref{eq:VN bound}, we derive
\begin{align}\label{eq:secret_key_bound1}
s_n&\geq n R_\text{asy}-\sqrt{n}\Delta_\text{AEP}(p_\text{EC}\epsilon_\text{s}^2/3,2^{2p})\notag\\+&\log_2(1-\epsilon^2_\text{s}/3)+2\log_2\sqrt{2}\epsilon_\text{h}
\end{align}
where we can use  the asymptotic secret key rate of Eq.~\eqref{eq:AsyRate}.
After a successful PE, the parties compute  $R_\text{asy}$ over a state $\tilde{\rho}_\text{wc}^n$ (instead  $\tilde{\rho}^n$), calculated with respect to the worst-case parameters given in Eq.~\eqref{eq:wcest} along with the worst case scenario entropy in Eq.~\eqref{eq:wcent}. 
As a result, Eq.~(\ref{eq:epsilon_def}) is replaced by the following
\begin{equation}\label{eq:security_one}
p_\text{EC}D(\tilde{\rho}_\text{wc}^n,\rho_\text{id})\leq \epsilon_\text{cor}+\epsilon_\text{h}+\epsilon_\text{s}.
\end{equation}
However, there is still the probability that the actual state is a bad state $\tilde{\rho}^n_\text{bad}$ with  probability $\tilde{\epsilon}_\text{PE}=2\epsilon_\text{PE}+\epsilon_\text{ent}$. On average, this is given by
\begin{equation}
\rho_\text{PE}=(1-\tilde{\epsilon}_\text{PE})\tilde{\rho}_\text{wc}^n+\tilde{\epsilon}_\text{PE}\tilde{\rho}^n_\text{bad}
\end{equation}
whose distance from the assumed worst-case state is
\begin{equation}\label{eq:security_two}
p_\text{EC}D(\rho_\text{PE},\tilde{\rho}_\text{wc}^n)   \leq p_\text{EC}\tilde{\epsilon}_\text{PE}.
\end{equation}

By using  Eqs.~(\ref{eq:security_one}) and (\ref{eq:security_two}),  together with the triangle inequality, we have that 
\begin{equation}\label{eq:compasable security}
p_\text{EC}D(\rho_\text{PE},\rho_\text{id})   \leq \epsilon_\text{cor}+\epsilon_\text{h}+\epsilon_\text{s}+p_\text{EC}(2\epsilon_\text{PE}+\epsilon_\text{ent}).
\end{equation}
Then the secret key length can be bounded by
\begin{align}\label{eq:secret_key_bound}
s_n&\geq n R_M-\sqrt{n}\Delta_\text{AEP}(p_\text{EC}\epsilon_\text{s}^2/3,2^{2p})\notag\\+&\log_2(1-\epsilon^2_\text{s}/3)+2\log_2\sqrt{2}\epsilon_\text{h},
\end{align}
where $R_M$ has been taken from Eq.~(\ref{eq:Mrate}). Finally, our previous specific analysis of the EC process allows us to connect $R_M$ with the practical rate $R^\text{EC}_M$ through the parameter $\hat{\beta}$ in Eq.~(\ref{eq:recEF}). By replacing the latter in the previous secret key bound and multiplying by the successful probability of a block $p_\text{EC}$ over the number of signals per block $N$, we obtain the composable secret key rate of Eq.~(\ref{rateCOMPO}).
 
Note that, although the concatenation of the quadratures may not be applied in practice, theoretically, it has to be considered for the calculation of the discretization parameter $|\mathcal{L}|$ included in the correction term $\Delta_\text{AEP}$. In fact, considering the proposed EC procedure, $|\mathcal{L}|$ takes the value $2p$ instead of $p$, compared with the case of the homodyne protocol~\cite{QKD_SIM}. In turn, this affects the compression needed to extract a secret key with length $s_n$.

\section{\label{app:entropy for calssical mapping}Classical data mapping and smooth-min entropy}
Let us assume a bidirectional mapping $X\leftrightarrow Z=f(X)$ where $X$ is a discrete random variable taking values $x$ in the alphabet $\mathcal{X}$ with probability $p_X$. Then, $Z$ takes values $z=f(x) \in \mathcal{Z}$ with probability $p_Z$. In fact, the probability function can absorb the action of $f$ such that  
\begin{equation}\label{eq:main_property}
 p_Z(z)=  p_Z(f(x))  =p_X(x).
\end{equation}
Therefore, the probabilities for the letters in $\mathcal{Y}$ are the same for the corresponding letter in $\mathcal{X}$.

We want to investigate what is the effect on $H^{\epsilon_s}_\text{min}$ of such a mapping, when it is applied to the classical system of the CQ state 
\begin{equation}
    \rho_{XE}=\sum_{x} p_X(x)| x \rangle_X \langle x | \otimes \rho_{E}(x).
\end{equation}
To do so we adapt the proof of~\cite[Prop.~6.20]{Tomamichel_book} for the state $\rho_{E}$ instead of $\rho_{AB}$. Thus we apply the isometry $U:U_X \otimes I_E$, with
$U_X : |x \rangle \mapsto |x \rangle_{X^{\prime}} |f(x)\rangle_Z$ being the Stinespring dilation of $f$ and $I_E$ the identity. As a result, we obtain the state
\begin{equation}
    \tau_{X^\prime Z E}=U \rho_{XE} U^\dagger.
\end{equation}
According to the invariance of the smooth min-entropy under isometries (see~\cite[Corollary~6.11]{Tomamichel_book}), we have the following relation
\begin{equation}\label{eq:isometry_invariance}
H_\text{min}^{\epsilon}(X|E)_\rho =H_\text{min}^{\epsilon}(X^\prime Z|E)_\tau.
\end{equation}
Furthermore, by using~\cite[Lemma~6.17]{Tomamichel_book}, we may write
\begin{equation}
    H_\text{min}^{\epsilon}(X^\prime Z|E)_\tau \geq  H_\text{min}^{\epsilon}( Z|E)_\tau
\end{equation}
for
\begin{equation}\label{eq:tracing_out classical subsystem}
    \tau_{ZE}=\sum_{x} p_X(x)| f(x)\rangle_Z \langle f(x) | \otimes \rho_{E}(x).
\end{equation}
From Eq.~\eqref{eq:isometry_invariance} and~\eqref{eq:tracing_out classical subsystem} we finally obtain
\begin{equation}\label{eq:one_system}
    H_\text{min}^{\epsilon}(X|E)_\rho \geq  H_\text{min}^{\epsilon}( Z|E)_\tau.
\end{equation}

Note that, in the same way, Eq.~\eqref{eq:one_system} can be extended to the case of two classical systems $X$ and $Y$ considering a Stinespring dilation $U_{XY}=U_XU_Y$ with $U_X : |x \rangle \mapsto |x \rangle_{X^{\prime}} |f(x)\rangle_Z$ and $U_Y : |y \rangle \mapsto |y \rangle_{Y^{\prime}} |f(y)\rangle_{Z'}$. Combining then Eq.~\eqref{eq:isometry_invariance} and~\eqref{eq:tracing_out classical subsystem} for the state 
\begin{align}
\rho_{XYE}=&\sum_{xy}p_{XY}(xy)|x\rangle_X\langle x|\notag \\
&\otimes|y\rangle_Y\langle y|\otimes  \rho_E(x,y),
\end{align}
one may write
\begin{equation}
    H_\text{min}^\epsilon(XY|E)_\rho\geq H_\text{min}^\epsilon(ZZ'|E)_\tau ,
\end{equation}
where
\begin{align}
\tau_{ZZ'E}=&\sum_{xy}p_{XY}(x,y)|f(x)\rangle_Z\langle f(x)|\notag\\
&\otimes |f(y)\rangle_{Z'}\langle f(y)| \otimes \rho_{E}(x,y).
\end{align}


\begin{thebibliography}{99}
\bibitem{entropic}For a function $f$ applied on a random variable $X$ we have
\begin{align}
H(X,f(X))&=H(X)+\underbrace{H(f(X)|X)}_{=0}\notag\\&=H(f(X))+H(X|f(X))
\end{align}
 The uncertainty for $f(X)$ given $X$ is vanishing which allows for the inequality 
\begin{equation}
    H(X)\geq H(f(X))\notag
\end{equation}
to be valid.
\end{thebibliography}
 \end{document}